\documentclass[11pt, a4paper]{article}
\usepackage[margin=1in]{geometry}
\usepackage{graphicx}
\usepackage{hyperref}
\usepackage{cite}
\usepackage{authblk}
\usepackage{siunitx}
\usepackage{xspace}
\usepackage{upgreek}
\usepackage{amsmath}
\usepackage{booktabs}
\usepackage{placeins}
\def\thebaroffset{0.18em}
\newcommand{\offsetoverline}[2][\thebaroffset]{\kern #1\overline{\kern -#1 #2}}%
\def\PB      {\ensuremath{B}\xspace}
\def\Pb      {\ensuremath{b}\xspace}
\def\B       {{\ensuremath{\PB}}\xspace}
\def\Bbar    {{\ensuremath{\offsetoverline{\PB}}}\xspace}
\def\bquarkbar      {{\ensuremath{\offsetoverline{\Pb}}}\xspace}
\def\Bb      {{\ensuremath{\Bbar}}\xspace}
\def\BorBbar {\kern \thebaroffset\optbar{\kern -\thebaroffset \PB}\xspace}
\def\Bz      {{\ensuremath{\B^0}}\xspace}
\def\Bzb     {{\ensuremath{\Bbar{}^0}}\xspace}
\def\Bs      {{\ensuremath{\B^0_s}}\xspace}

\begin{document}

\title{
Inclusive Flavour Tagging at LHCb\\[4pt]
\large \textit{Presented at the 32nd International Symposium on Lepton Photon Interactions\\
at High Energies, Madison, Wisconsin, USA, August 25–29, 2025}
}

\author[1]{Jonah Blank \thanks{On behalf of the LHCb collaboration.}}
\affil[1]{Fakultät Physik, Technische Universität Dortmund, Dortmund, Germany}

\date{} 
\maketitle

\begin{center}
\textit{This proceedings article summarizes work originally presented in Ref.~\cite{originalpaper} and is submitted with permission from the original authors.}
\end{center}

\begin{abstract}
\noindent A new algorithm based on a deep neural network, DeepSets, for tagging the production flavour of neutral $B^0$ and $B^0_s$ mesons in proton-proton collisions is presented. Exploiting a comprehensive set of tracks associated with the hadronization process, the algorithm is calibrated on data collected by the LHCb experiment at a centre-of-mass energy of \SI{13}{\tera\eV}. This inclusive approach enhances the flavour tagging performance beyond the established same-side and opposite-side tagging methods. The observed gains in tagging power of $35\%$ for $B^0$ mesons and $20\%$ for $B_s^0$ mesons relative to the combined performance of the existing LHCb flavour-tagging algorithms offer significant benefits for precision measurements of $C\!P$ violation and mixing in the neutral $B$ meson systems.
\end{abstract}

\section{Introduction}
Precise knowledge of the production flavour of neutral $B$ mesons is crucial for time-dependent measurements of $B^0_{(s)}$--$\bar{B}^0_{(s)}$ oscillations and CP-violating observables. At the LHCb experiment, this information is inferred using flavour-tagging algorithms that analyse particles in the periphery of the signal decay to predict the $b$-quark content of the initial meson.\\
The current LHCb tagging strategy relies on two complementary approaches: opposite-side (OS) taggers, which use decay products of the other $b$ produced in the $b\overline{b}$ pair production, and same-side (SS) taggers, which exploit particles produced in association with the signal $B$ meson during hadronisation. Despite their success, the overall tagging power remains limited by the selection needed to determine the correct track in the event for tagging purposes. With an increased instantaneous luminosity due to the implementation of a purely software based trigger for the LHCb Upgrade I~\cite{UpgradeI} and future plans for the HL-LHC, this challenge will get even harder leading to the idea of a more inclusive tagging approach to mitigate these issues.\\
In Ref.~\cite{originalpaper}, a novel \emph{Inclusive Flavour Tagger} (IFT) is introduced, based on a deep neural network using the DeepSets architecture, incorporating information from all tracks in the event. It is trained separately on simulation and calibrated on data collected by the LHCb experiment during 2016, 2017 and 2018 at $\SI{13}{\tera\eV}$.\\
This contribution provides a concise summary of the IFT approach, its calibration strategy, and the key performance results, as presented at the 32nd International Symposium on Lepton Photon Interactions. 

\section{Methodology and training}
The classical SS and OS taggers rely on specialised subsets of tracks associated with particular physics processes. While these taggers incorporate physics knowledge into the decision, their predefined track selections lead to ambiguities and inevitably discard correlations present in the full event. In contrast, the IFT uses all reconstructed tracks in the event, thereby retaining this information. A challenge of this inclusive strategy is the varying number of tracks per event. The DeepSet architecture, consisting of two Deep Neural Networks (DNNs) $\phi$ and $\rho$ handles a variable input size. First, the per-track feature vectors are processed independently by $\phi$. The resulting latent vectors are then aggregated into a single permutation-invariant representation, which is passed to $\rho$, to produce the tagging output. The network $\rho$ comprises an input layer, three hidden layers, and an output layer. Binary Cross Entropy is used as the loss function. In total, the model has 3980 free parameters to be trained. The model is trained using the binary cross-entropy loss and contains a total of 3980 trainable parameters. A rectified linear unit activation function is used throughout, except for the final sigmoid output, which ensures a probabilistic prediction $y \in[0,1]$.\\
Separate classifiers are trained for \Bz and \Bs due to the different SS fragmentation of the $d$ and $s$ quark, and for the data-taking periods (2016-2018) to reflect changes in the trigger configuration. For the training, simulated samples of $B^0\to J/\psi K^*(892)^0$ and $B_s^0\to J/\psi\phi(1020)$ are used, while the calibration is performed on background-subtracted real data from the flavour-specific decays $B^0\to J/\psi K^+\pi^-$ and $B_s^0\to D_s^-\pi^+$ are used.
The performance is validated on the golden mode $B^0\to J/\psi K_S$, used to measure $\sin{2\beta}$, and on $B_s^0\to J/\psi K^+K^-$, which is sensitive to the weak mixing phase $\phi_s$.\\
Before training the IFT, a multiclass Boosted Decision Tree~(BDT) is trained using kinematic information to categorise tracks into SS fragmentation, OS fragmentation, OS decay or unrelated. Its outputs scores are included as an additional input to the DeepSet-IFT, together with the BDT input features. The predictions of the classical taggers are also provided as optional inputs. Although the IFT already outperforms the classical taggers when trained solely on low-level inputs, including these higher-level features yields a further improvement in the effective tagging power,
\[
\varepsilon_\text{eff} = \varepsilon_\text{tag}\langle D^2\rangle,
\]
Here, $\varepsilon_\text{tag}$ denotes the tagging efficiency and $\langle D^2\rangle$ is the average squared dilution. For classical taggers, the efficiency is limited by the need to correctly select a specific track, whereas for the IFT only events with mistag of exactly $\eta=0.5$ are excluded, leading to a much higher $\varepsilon_\text{tag}$ . The dilution, related to the mistag probability through $\omega = \frac{1-D}{2}$, is determined using a maximum-likelihood fit to the decay-time distribution. An increase in $\varepsilon_\text{eff}$ directly translates to a reduction of the statistic uncertainty, $\sigma_{stat}\propto\frac{1}{\sqrt{\varepsilon_\text{eff}N}}$, where $N$ is the number of selected events.\\
The tagging decision $d$ and per-event mistag estimate $\eta$ are defined as
\[
(d,\eta) = \begin{cases}
    (\;\;\;1,\,1-y) \text{\;\;\;\; for } y > 0.5,\\
    (-1, \;\;\;\;\;\;\;y) \text{\;\;\;\; for } y< 0.5,
\end{cases} 
\]
where $d=\pm1$ corresponds to a \Bz or \Bzb. Due to residual differences between simulation and real data, $\eta$ must be calibrated using a linear fit to calibration data:
\[
\omega^{\B/\Bb}(\eta) = p_0 \pm p_1\cdot(\eta - \offsetoverline{\eta}) \pm \frac{1}{2}\Delta\omega,
\]
where $\offsetoverline{\eta}$ is the average mistag in the training sample, and
\[
\Delta\omega = \Delta p_0 \pm \Delta p_1 \cdot (\eta - \offsetoverline{\eta})
\]
accounts for particle/anti-particle differences when the optimal threshold satisfies $\eta_\text{opt}\neq 0.5$.
\section{Performance results of the DeepSet-IFT}
Since the classical taggers only operate on specific physics cases, their outputs need to be combined to allow a fair comparison with the IFT. First, the individual SS and OS taggers are combined into a single SS and an OS decision by computing the probability that the signal candidate contained a \Pb or a \bquarkbar during production:
\[
P(\Pb) = \frac{p(\Pb)}{p(\Pb) + p(\bquarkbar)}\text{, }P(\bquarkbar) = 1 - P(\Pb),
\]
where
\begin{align*}
    p(\Pb) &= \prod_i\left(\frac{1+d_i}{2}-d_i(1-\eta_i)\right) \text{ and}\\
    p(\bquarkbar) &= \prod_i\left(\frac{1-d_i}{2}+d_i(1-\eta_i)\right)
\end{align*}
assuming the individual taggers to be uncorrelated.
From these probabilities, the combined tagging decision and combined mistag estimate are defined as
\begin{align}
    d_\text{comb} &= \mathrm{sign}(P(\bquarkbar) - P(\Pb))\text{ and }\\
    \eta_\text{comb} &= 1 - \mathrm{max}(P(\Pb),P(\bquarkbar)).
\end{align}
The same procedure is applied to combine the SS and OS results into a single tagging decision and mistag probability estimate.
\subsection{The \Bz-IFT}
As for the IFT, the mistag probability of the SS and OS combination is calibrated on data using $\Bz\to J/\psi K^+\pi^-$ decays. The resulting performance for the different data-taking periods is summarised in Table~\ref{tab:per_b0}.
An improvement with the IFT is observed in both $\varepsilon_\text{tag}$ and $\langle D^2\rangle$, leading to a consistent enhancement in tagging power of about $35\%$ compared to the classical SS and OS combination across all datasets.
\begin{table}[h!]
\centering
\caption{Performance with statistical uncertainty for the classical taggers and the IFT on real $\Bz\to J/\psi K^+\pi^-$ decay candidates.}\label{tab:per_b0}
\begin{tabular}{l l c c c}
\toprule
Year & Tagger & $\varepsilon_{\text{tag}}(\%)$ & $\langle D^2 \rangle$ & $\varepsilon_{\text{tag}}\langle D^2\rangle (\%)$ \\
\midrule
2016 & SS and OS & $87.030 \pm 0.058$ & $0.0466 \pm 0.0010$ & $4.05 \pm 0.09$ \\
     & IFT        & $95.789 \pm 0.035$ & $0.0582 \pm 0.0011$ & $5.58 \pm 0.10$ \\
\midrule
2017 & SS and OS & $87.208 \pm 0.058$ & $0.0485 \pm 0.0010$ & $4.23 \pm 0.09$ \\
     & IFT        & $95.024 \pm 0.038$ & $0.0617 \pm 0.0011$ & $5.87 \pm 0.10$ \\
\midrule
2018 & SS and OS & $87.631 \pm 0.052$ & $0.0493 \pm 0.0010$ & $4.32 \pm 0.09$ \\
     & IFT        & $95.172 \pm 0.034$ & $0.0615 \pm 0.0010$ & $5.85 \pm 0.10$ \\
\bottomrule
\end{tabular}
\end{table}\\\noindent
The portability of the \Bz-IFT is tested using the decay $B^0\to J/\psi K_S$ that was not used during training. For a proper mistag calibration, events from $\Bz\to J/\psi K^+\pi^-$ are reweighted such that their kinematic and multiplicity distributions match those of $B^0\to J/\psi K_S$. The corresponding performance numbers are given in Table~\ref{tab:val_b0}. Here, the increase in tagging power is even larger, around $65\%$, which can be attributed a suboptimal selection leading to a significantly lower $\varepsilon_\text{tag}$. It is worth noting that in the most recent $\sin{(2\beta)}$ measurement, Ref.~\cite{sin2beta} the tagging efficiency is given as $(85.34\pm0.05)\%$, while the dilution is similar to what is observed her for the SS and OS combination. Assuming those values the improvement would be in consistent with that obtained in the $\Bz\to J/\psi K^+\pi^-$ calibration channel.
\begin{table}[ht]
\centering
\caption{Performance with statistical uncertainty for the classical taggers and the IFT on real $\Bz\to J/\psi K_S$ decay candidates.}\label{tab:val_b0}
\begin{tabular}{l l c c c}
\toprule
Year & Tagger & $\varepsilon_{\text{tag}}(\%)$ & $\langle D^2 \rangle$ & $\varepsilon_{\text{tag}}\langle D^2\rangle (\%)$ \\
\midrule
2016 & SS and OS & $69.18 \pm 0.35$ & $0.0473 \pm 0.0011$ & $3.27 \pm 0.08$ \\
     & IFT        & $95.42 \pm 0.16$ & $0.0563 \pm 0.0012$ & $5.37 \pm 0.12$ \\
\midrule
2017 & SS and OS & $68.65 \pm 0.34$ & $0.0482 \pm 0.0011$ & $3.31 \pm 0.08$ \\
     & IFT        & $98.53 \pm 0.09$ & $0.0571 \pm 0.0012$ & $5.63 \pm 0.12$ \\
\midrule
2018 & SS and OS & $69.58 \pm 0.31$ & $0.0487 \pm 0.0010$ & $3.38 \pm 0.07$ \\
     & IFT        & $96.41 \pm 0.13$ & $0.0573 \pm 0.0011$ & $5.52 \pm 0.11$ \\
\bottomrule
\end{tabular}
\end{table}
\FloatBarrier
\subsection{The \Bs-IFT}
Analogously, the performance of the \Bs-IFT is evaluated on data using the calibration channel $B_s^0\to D_s^-\pi^+$, while $B_s^0\to J/\psi K^+K^-$ is used to validate the portability.
The corresponding values are shown in Tables~\ref{tab:per_Bs} and \ref{tab:val_Bs}.
While the absolute tagging power differs between these decay modes, a consistent improvement of about $20\%$ is observed in both channels. Furthermore, a blinded fit to $B_s^0\to J/\psi K^+K^-$ following the strategy in Ref.~\cite{phi_s} is performed to assess the impact on the precision of the $C\!P$ violating phase $\phi_s$. The statistic uncertainty is found to decrease by about $10\%$, consistent with the expected behaviour given the $20\%$ increase in $\varepsilon_\text{eff}$.
\begin{table}[ht]
\centering
\caption{Performance with statistical uncertainty for the classical taggers and the IFT on real $B_s^0 \to D_s^- \pi^+$ decay candidates.}\label{tab:per_Bs}
\begin{tabular}{l l c c c}
\toprule
Year & Tagger & $\varepsilon_{\text{tag}}(\%)$ & $\langle D^2 \rangle$ & $\varepsilon_{\text{tag}}\langle D^2\rangle (\%)$ \\
\midrule
2016 & SS and OS & $73.81 \pm 0.16$ & $0.0877 \pm 0.0054$ & $6.47 \pm 0.40$ \\
     & IFT        & $87.97 \pm 0.13$ & $0.0882 \pm 0.0050$ & $7.76 \pm 0.44$ \\
\midrule
2017 & SS and OS & $79.05 \pm 0.15$ & $0.0758 \pm 0.0043$ & $5.99 \pm 0.34$ \\
     & IFT        & $96.87 \pm 0.06$ & $0.0747 \pm 0.0038$ & $7.23 \pm 0.37$ \\
\midrule
2018 & SS and OS & $74.34 \pm 0.15$ & $0.0819 \pm 0.0042$ & $6.09 \pm 0.31$ \\
     & IFT        & $95.49 \pm 0.07$ & $0.0760 \pm 0.0036$ & $7.25 \pm 0.34$ \\
\bottomrule
\end{tabular}
\end{table}
\begin{table}[ht]
\centering
\caption{Performance with statistical uncertainty for the classical taggers and the IFT on real $B_s^0 \to J/\psi K^+K^-$ decay candidates.}\label{tab:val_Bs}
\begin{tabular}{l l c c c}
\toprule
Year & Tagger & $\varepsilon_{\text{tag}}(\%)$ & $\langle D^2 \rangle$ & $\varepsilon_{\text{tag}}\langle D^2\rangle (\%)$ \\
\midrule
2016 & SS and OS & $74.91 \pm 0.14$ & $0.0623 \pm 0.0039$ & $4.67 \pm 0.29$ \\
     & IFT        & $91.18 \pm 0.09$ & $0.0609 \pm 0.0035$ & $5.55 \pm 0.32$ \\
\midrule
2017 & SS and OS & $76.18 \pm 0.14$ & $0.0582 \pm 0.0034$ & $4.44 \pm 0.26$ \\
     & IFT        & $93.18 \pm 0.08$ & $0.0578 \pm 0.0030$ & $5.38 \pm 0.28$ \\
\midrule
2018 & SS and OS & $72.31 \pm 0.13$ & $0.0628 \pm 0.0033$ & $4.54 \pm 0.24$ \\
     & IFT        & $90.32 \pm 0.09$ & $0.0604 \pm 0.0029$ & $5.46 \pm 0.26$ \\
\bottomrule
\end{tabular}
\end{table}
\subsection{Uncertainty studies}
Using a tagger on a decay channel it was not trained on introduces a systematic uncertainty because the difference in kinematics and multiplicity can affect the calibration. In some analysis, such as the measurement of $\sin{(2\beta)}$, this contribution may even be dominant. 
To quantify this effect, the calibration parameters obtained from simulated and reweighted $\Bz\to J/\psi K^*$($B_s^0\to D_s^-\pi^+$) events are compared with those obtained directly from $B^0\to J/\psi K_S$($B_s^0\to J/\psi\phi(1020)$) for \Bz(\Bs). For the \Bz taggers, the differences are larger than the statistical uncertainty, as expected, but similar for both the SS and OS combination and the IFT. Likewise, for the \Bs taggers the size of the effect is similar for both the classical taggers and the IFT, and remains subdominant relative to the statistical uncertainties. This indicates that porting the IFT to other decay modes does not introduce a larger systematic uncertainty than the currently used taggers.

\section{Summary and Outlook}
A new inclusive flavour tagger based on the DeepSets architecture has been developed and its performance evaluated using Run~2 data collected by the LHCb experiment. By exploiting information from all reconstructed tracks, the IFT achieves a significantly higher tagging power than the classical SS and OS taggers. For \Bz mesons, improvements of up to $35\%$ in the effective tagging power are observed, while for \Bs mesons an enhancement of around $20\%$ is obtained. These gains translate into a corresponding reduction of the statistical uncertainty in time-dependent $C\!P$ violation measurements, as explicitly demonstrated in a blinded study of the $B_s^0\to J/\psi K^+K^-$ decay.\\
Systematic uncertainties associated with porting the tagger to decay modes with different kinematic and multiplicity distributions have been assessed. The size of these effects is comparable to that of the currently used taggers and does therefore not represent a limiting factor for the IFT.\\
The demonstrated performance shows that inclusive flavour tagging is a viable and advantageous alternative for LHCb Run~2 analyses. While the IFT is now included into the LHCb framework allowing currently on-going analyses to benefit from increased statistics, an effort is made to also develop an inclusive tagger for the more challenging environment of Run~3 and beyond, including different architectures, such as transformers, to achieve optimal performance.


\end{document}